\documentclass[conference]{IEEEtran}
\usepackage{graphicx}
\usepackage{subfigure}
\usepackage{float}
\usepackage{amsmath}
\usepackage{amsfonts,amssymb}
\usepackage{dsfont}
\usepackage{algpseudocode}
\usepackage{makecell}
\usepackage{cite}
\usepackage{multirow}
\usepackage{multicol}
\usepackage{booktabs}
\usepackage[ruled,vlined, linesnumbered,lined,boxed,commentsnumbered,ruled,longend,noend]{algorithm2e}
\usepackage{xcolor,soul}
\usepackage{comment}
\usepackage{bbm}

\SetKwInput{KwInput}{Input}                % Set the Input
\SetKwInput{KwOutput}{Output}              % set the Output

\usepackage{fancyhdr}
\fancypagestyle{ms}{
   \fancyhf{}
   
   \fancyhead[C]{Reprinting or republishing this material for the purpose of advertising or promotion, creating new collective works, reselling or redistributing to servers or lists, or using any copyrighted component in other works must adhere to IEEE policy.}
}
\makeatletter
\let\ps@IEEEtitlepagestyle\ps@ms
\makeatother

\begin{document}
\title{Fairness-Utilization Trade-off in Wireless Networks with Explainable Kolmogorov-Arnold Networks}

\author{
    \IEEEauthorblockN{
        Masoud Shokrnezhad\textsuperscript{1}, Hamidreza Mazandarani\textsuperscript{2}, and Tarik Taleb\textsuperscript{2} \\
    }
    \IEEEauthorblockA{
       % \textit{1 IEEE Member\\
        \textsuperscript{1} \textit{Oulu University, Oulu, Finland; masoud.shokrnezhad@oulu.fi} \\
        \textsuperscript{2} \textit{Ruhr University Bochum, Bochum, Germany; hr.mazandarani@ieee.org, tarik.taleb@rub.de}
    }
}

\maketitle

\begin{abstract}
The effective distribution of user transmit powers is essential for the significant advancements that the emergence of 6G wireless networks brings. In recent studies, Deep Neural Networks (DNNs) have been employed to address this challenge. However, these methods frequently encounter issues regarding fairness and computational inefficiency when making decisions, rendering them unsuitable for future dynamic services that depend heavily on the participation of each individual user. To address this gap, this paper focuses on the challenge of transmit power allocation in wireless networks, aiming to optimize $\alpha$-fairness to balance network utilization and user equity. We introduce a novel approach utilizing Kolmogorov-Arnold Networks (KANs), a class of machine learning models that offer low inference costs compared to traditional DNNs through superior explainability. The study provides a comprehensive problem formulation, establishing the NP-hardness of the power allocation problem. Then, two algorithms are proposed for dataset generation and decentralized KAN training, offering a flexible framework for achieving various fairness objectives in dynamic 6G environments. Extensive numerical simulations demonstrate the effectiveness of our approach in terms of fairness and inference cost. The results underscore the potential of KANs to overcome the limitations of existing DNN-based methods, particularly in scenarios that demand rapid adaptation and fairness.
\end{abstract}


\section{Introduction}
The advent of 6G wireless networks heralds a new era of connectivity, promising to revolutionize sectors such as healthcare, education, logistics, and transportation \cite{shokrnezhad2024semantic}. These next-generation networks are poised to deliver unprecedented capabilities, including ultra-high data rates, massive device connectivity, and adaptive responses to highly dynamic environments \cite{yu2023toward}. Central to realizing these advancements is the efficient allocation of user transmit powers, a critical factor that directly influences network performance, user experience, and energy efficiency. In recent years, the complexity of this challenge has led researchers to explore innovative solutions leveraging Machine Learning (ML) techniques, with a particular focus on Deep Neural Networks (DNNs).

Among the studies addressing the transmit power allocation problem in this rapidly evolving field, several notable approaches stand out. Nasir \textit{et al.} \cite{Nasir8792117} and Sheu \textit{et al.} \cite{Sheu10224533} employed Deep Q-Learning (DQL) to maximize the sum data rate of users, utilizing channel information as input. Li \textit{et al.} \cite{Li9770401} extended the same approach to a distributed setting. Jamous \textit{et al.} \cite{Jamous10017530} applied DQL to optimize transmission energy efficiency. Zhang \textit{et al.} \cite{Zhang8861118} innovated by using convolutional DNNs with users' geographical information to maximize aggregate data rates. In a different approach, Zhang \textit{et al.} \cite{Zhang9046301} implemented Proximal Policy Optimization (PPO) with signal strength inputs to ensure predefined Signal-to-Interference-plus-Noise Ratio (SINR) thresholds. Huang \textit{et al.} \cite{Huang9850358} also utilized PPO, focusing on maximizing the sum of data rates.

While existing DNN-based methods have demonstrated considerable performance, they face significant challenges in two key areas: balancing network utilization with fairness, and achieving computational efficiency during inference. Most of the existing studies have primarily focused on system-wide performance indicators, such as aggregate data rates, often at the expense of equitable resource allocation among individual users. This oversight becomes particularly critical in the context of future services, where semantic-aware communication is expected, and ensuring fair participation for each user is essential to maintain the quality and diversity of outcomes, thereby mitigating potential biases from specific sources \cite{mazandarani2024semantic}. Furthermore, the DNN-based techniques prevalent in the literature are predominantly black-box models, necessitating complex computations for each inference. This computational intensity often results in prolonged inference times \cite{khan2023explainable}. Such inefficiency is particularly problematic in the dynamic environments anticipated for 6G services, where rapid adaptation to changing environmental conditions is paramount. 

To address the gaps in existing research, this paper focuses on investigating the power allocation problem with the objective of optimizing $\alpha$-fairness. The $\alpha$-fairness metric offers a versatile framework for balancing the trade-off between fairness and utilization in resource allocation. By modulating $\alpha$, we can achieve various fairness objectives, providing a flexible approach suitable for dynamic future services. To tackle this problem, we employ a novel class of machine learning models known as Kolmogorov-Arnold Networks (KANs), which have been proposed as an alternative to conventional DNNs \cite{liu2024kan}. KANs are designed to approximate continuous multivariate functions using learnable activation functions within a relatively simple architecture, offering improved generalization capabilities. The reliance on these functions renders KANs fully explainable, significantly reducing the computational overhead typically associated with inference. This characteristic makes KANs particularly well-suited for time-sensitive and resource-constrained environments, offering an attractive solution for next-generation communication systems.

The remainder of this paper is structured as follows. Section \ref{s_prb_stt} presents the system model, provides a comprehensive problem formulation, and proves the NP-hardness of the considered problem. Section \ref{s_prp_slt} elucidates the proposed KAN-based solution, encompassing its fundamental principles, as well as the proposed dataset generation and decentralized training algorithms. In Section \ref{s_sim}, we present and analyze numerical results, with a particular focus on evaluating the efficiency of the proposed solution in terms of fairness and inference cost. Finally, Section \ref{s_con} concludes the paper with a summary of our findings and closing remarks on the implications and potential future directions of this research.

\section{Problem Statement}\label{s_prb_stt}

\subsection{System Model}\label{ss_sys_mdl}
\begin{figure}[!t]
    \centerline{\includegraphics[width=3.5in]{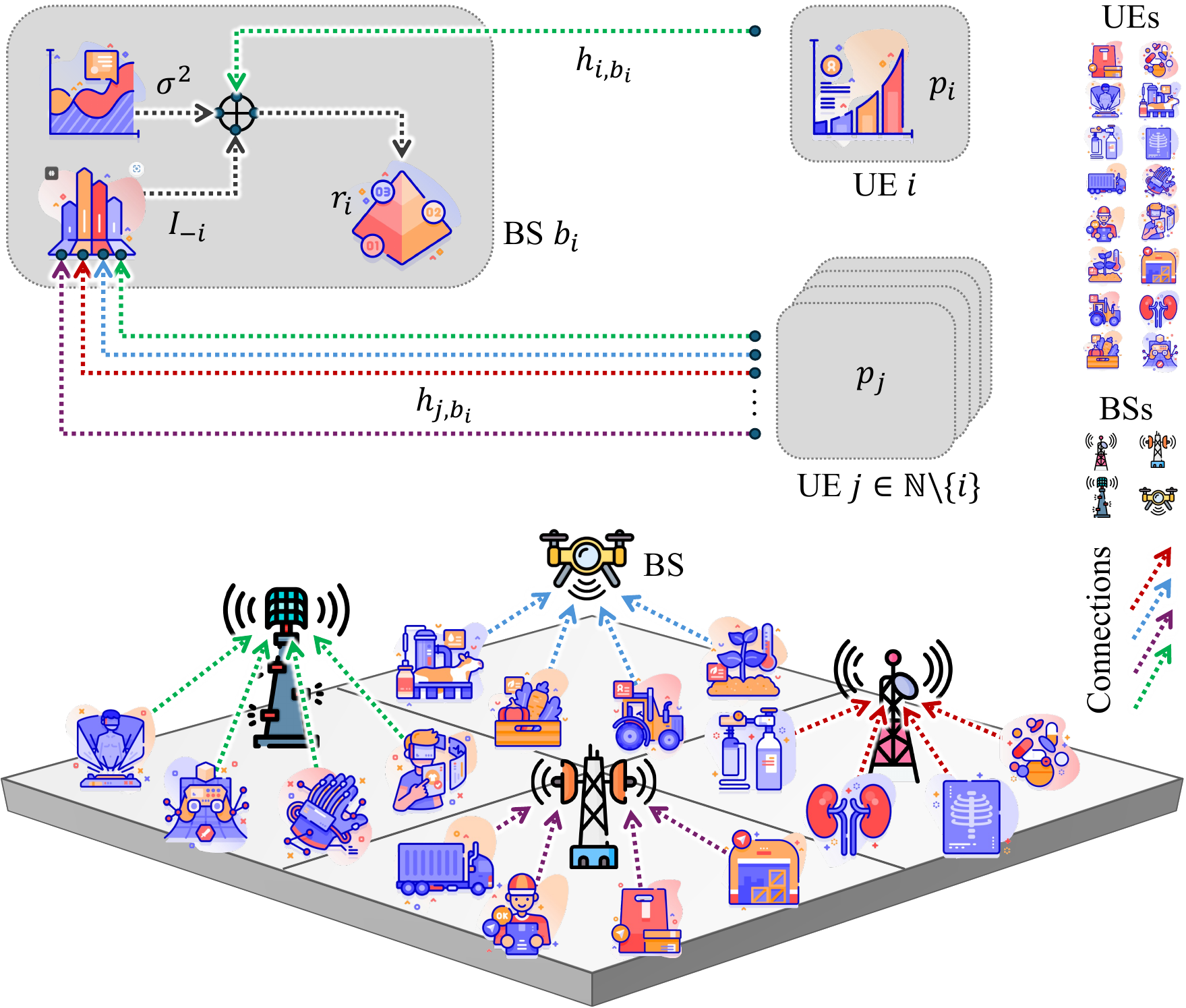}}
    \caption{The system model.}
    \label{fig1}
\end{figure}

This paper focuses on a wireless network, comprising a set of Base Stations (BSs) and User Equipment (UE), as illustrated in Fig. \ref{fig1}. The network is characterized by $\mathcal{B}$ BSs, denoted as $\mathbb{B}$, and $\mathcal{N}$ intelligent UE, represented by $\mathbb{N}$. We exclusively consider uplink transmissions, and the data rate for each UE $i$ is calculated using the Shannon-Hartley theorem. This fundamental theorem in information theory defines the channel capacity, and consequently, the maximum achievable data rate, as follows:
\begin{align} \label{rates}
    r_{i} &= \log_{2}{ \Big( 1 + \gamma_{i} \big) } \notag \\
    &= \log_{2}{ \big( 1 + \frac{p_i \times h_{i,{b}_{i}}}{\mathcal{I}_{-i} + {\sigma}^{2}} \Big) } \notag \\
    &= \log_{2}{ \big( 1 + \frac{p_i \times h_{i,{b}_{i}}}{\sum\limits_{j \in \mathbb{N} \setminus \{i \} }^{}{p_j \times h_{j,{b}_{i}}} + {\sigma}^{2}} \Big) }.
\end{align}
In this formulation, the following variables and parameters are defined:
\begin{itemize}
    \item ${b}_{i}$ represents the BS associated with UE $i$. For the purposes of this study, it is assumed that UEs are uniformly distributed among BSs such that the average distance between each UE and its serving BS is minimized.
    \item $p_i$ denotes the transmit power of UE $i$, and $\mathbb{P}$ represents the vector encompassing the transmit power of all UEs.
    \item $h_{i, b_i}$ signifies the path gain between UE $i$ and BS ${b}_{i}$.
    \item ${\sigma}^{2}$ represents the noise power.
    \item $\gamma_{i}$ denotes the SINR of UE $i$ at BS ${b}_{i}$.
    \item $\mathcal{I}_{-i}$ quantifies the interference power at BS ${b}_{i}$, which is equivalent to the sum of received power from all UEs except UE $i$ at BS ${b}_{i}$.
\end{itemize}

It is important to note that this paper assumes the implementation of a universal frequency reuse strategy, characterized by a reuse factor of one. This design choice is primarily motivated by the scarcity and high cost of spectrum resources, which necessitates efficient spectrum utilization. The implementation of this approach aligns with the principles established in Fifth-Generation New Radio (5G-NR) specifications. As we look towards the development of 6G wireless networks, this one-to-one reuse factor is anticipated to play an increasingly critical role. It is expected to be instrumental in achieving the ultra-high spectrum efficiency required to meet the stringent performance targets of future wireless communication systems. This approach provides a foundation for advanced techniques such as coordinated multipoint transmission and reception, which are crucial for realizing the full potential of next-generation wireless networks.

\subsection{Problem Formulation}\label{ss_prb_frm}

\begin{figure*}[!t]
    \centerline{\includegraphics[width=7in]{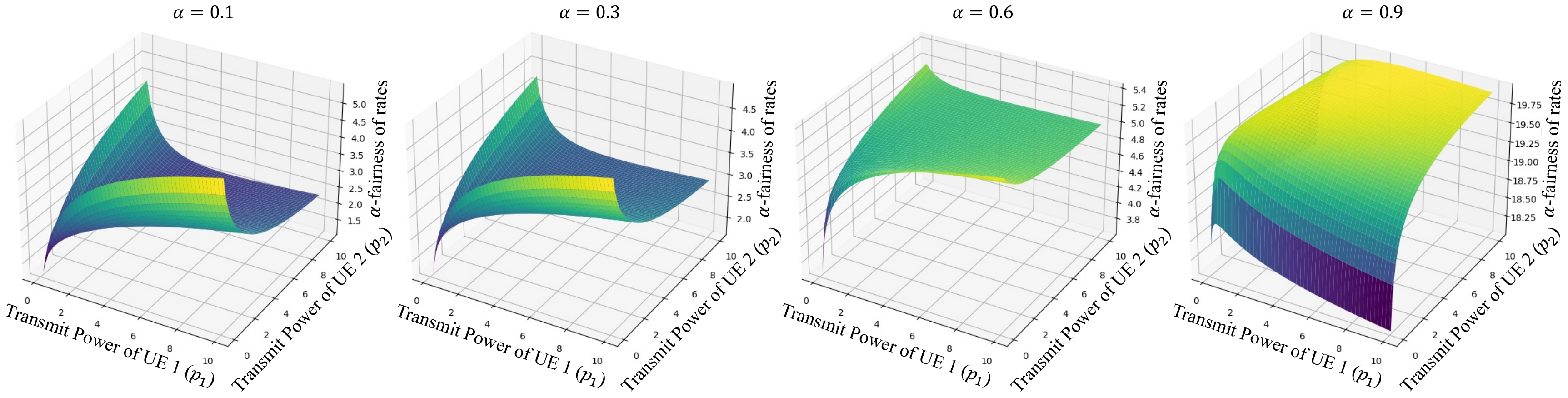}}
    \vspace{-10pt}
    \caption{The $\alpha$-fairness of data rates for a two-user network for various transmit powers. The subplots show $\alpha$-fairness for $\alpha$ values of $0.1$, $0.3$, $0.6$ and $0.9$, respectively. Assuming that there is just one base station, channel gains are set to $0.8$ for UE 1 and $0.4$ for UE 2, representing asymmetric channel conditions. $\sigma^2$ is fixed at $0.1 W$. Transmit powers for both users range from $0.1 W$ to $10 W$. The color gradient represents the magnitude of $\alpha$-fairness, with warmer colors indicating higher utility values.}
    \label{fig2}
    \vspace{-10pt}
\end{figure*}

In this study, we aim to optimize the network performance by maximizing the $\alpha$-fairness of UE data rates, that is:
\begin{align}\label{eq_frn}
    & \mathcal{F}_{\alpha} = 
    \begin{cases} 
        \sum\limits_{i \in \mathbb{N}}{ log(r_{i})} & \text{if } \alpha=1 \\
        (1 - \alpha)^{-1} \times \sum\limits_{i \in \mathbb{N}}{ {(r_{i})}^{1-\alpha}} & \text{if } \alpha \neq 1
    \end{cases}
\end{align}
 The trade-off between equitable resource distribution and overall system utilization is adjustable by balancing $\alpha$ in $\alpha$-fairness.
 %When $\alpha = 1$, the objective function becomes the sum of logarithms of the rates, equivalent to proportional fairness, balancing system rate and user fairness. For $\alpha \neq 1$, the function takes a more general form, enabling a spectrum of fairness criteria. 
 As depicted in Fig. \ref{fig2}, when $\alpha \to 0$, optimizing \eqref{eq_frn} becomes equivalent to maximizing the sum of data rates, commonly known as utilitarian fairness or sum-rate maximization. In this scenario, the objective is to maximize the total data rate, with minimal consideration for its distribution among users. Although this method is generally less complex, it can result in significant inequality. Conversely, when $\alpha \to \infty$, optimizing \eqref{eq_frn} shifts to maximizing the minimum UE's data rate, known as max-min fairness. Max-min fairness ensures that the transmit power allocations are adjusted to enhance the allocation received by the UE with the worst channel condition, thereby achieving the highest level of equality among users, albeit potentially at the expense of overall complexity. Given this, we formulate our $\alpha$-Fairness Power Allocation Problem ($\alpha$-FPAP) as shown in \eqref{eq_prb}.
\begin{align}\label{eq_prb}
    \alpha\text{-FPAP:} \quad & \max_{\mathbb{P}}{\mathcal{F}_{\alpha}} \quad \textit{s.t. Rate Constraints} \ \eqref{rates} \\
    & \textit{where} \quad \mathcal{P}_{\text{min}} \leq p_{i} \leq \mathcal{P}_{\text{max}} \quad \forall i \in \mathcal{N} \notag
\end{align}
Here, we seek to maximize $\alpha$-fairness by adjusting the transmit power of UEs subject to both rate and transmit power constraints, ensuring that the transmit power remains within the feasible range defined by \( \mathcal{P}_{\text{min}} \) and \( \mathcal{P}_{\text{max}} \).

\subsection{Complexity Analysis}\label{ss_cmp_anl}
The $\alpha$-FPAP intuitively appears NP-hard due to several compelling reasons. With \(\mathcal{N}\) UEs and \(\mathcal{B}\) BSs, the number of possible allocations for transmit powers exhibits exponential growth. Each UE's power setting influences not only its own data rate but also potentially impacts the rates of all other UEs due to mutual interference. The data rate for each UE is characterized by a non-linear function dependent on its own power and the interference received from other UEs, and the incorporation of $\alpha$-fairness introduces additional non-linearity into the objective function. Small perturbations in channel conditions or user locations can result in significant shifts in the optimal power allocation. The $\alpha$ parameter introduces further complexity, as varying $\alpha$ values can produce vastly different optimal allocations. These characteristics collectively result in a multi-dimensional solution space abundant with numerous local optima, rendering the identification of the global optimum particularly challenging.

Beyond intuitive reasoning, the NP-hardness of the $\alpha$-FPAP requires a formal proof. To establish this, we need to demonstrate that a well-known NP-hard problem can be reduced to a special case of the $\alpha$-FPAP in polynomial time. We choose the Maximum Independent Set Problem (MISP) for this reduction. Given an undirected graph \(\mathcal{G} = (\mathbb{V}, \mathbb{E})\) where \(\mathbb{V} = \{v_i \mid i \in \mathbb{N}\}\) and \(\mathbb{E} \subseteq \{(v_i, v_j) \mid i, j \in \mathbb{N}\}\), the MISP seeks to find the largest subset of vertices such that no two vertices in the subset are connected by an edge. The reduction proceeds as follows:
\begin{enumerate}
    \item Set $\alpha =  1$ (proportional fairness). 
    \item Set the noise power to a small value $\epsilon > 0$.
    \item Suppose that each vertex $v_i$ in $\mathcal{G}$ is equivalent with a UE-BS pair (UE $i$, $b_i$). Set channel gain $h_{i, b_i} = 1$ for each UE $i$ to its BS $b_i$. 
    \item For each pair ($v_i$, $v_j$) in $\mathcal{G}$, establish an edge if $h_{i, b_j} > \epsilon$ or $h_{j, b_i} > \epsilon$. Then, set channel gains $h_{i, b_j} = h_{j, b_i} = \mathcal{M}$, where $\mathcal{M}$ is a large positive value.
    \item Set all other channel gains to $0$.
\end{enumerate}

In this setup, the optimal solution to the $\alpha$-FPAP will allocate high power to UEs that correspond to a maximum independent set in \(\mathcal{G}\) (i.e., UEs whose corresponding vertices in \(\mathcal{G}\) are not connected by an edge), and low or negligible power to others. In other words, a maximum independent set in \(\mathcal{G}\) corresponds to the set of UEs that can transmit at high power without causing significant interference. Consequently, if the special version of the $\alpha$-FPAP (with \(\alpha = 1\) and specific channel gain settings) could be solved in polynomial time, the MISP could also be solved in polynomial time. Considering that the reduction is polynomial relative to the size of the input graph \(\mathcal{G}\) and the MISP is NP-hard, this signifies that the special version of the $\alpha$-FPAP is also NP-hard. Since this special version is at least as difficult as the general problem, the proof is complete, and the $\alpha$-FPAP is proved to be an NP-hard problem\footnote{This proof establishes NP-hardness for the decision version of the problem. The optimization version (finding the actual power allocation) is at least as hard as the decision version.}.

\section{Proposed Solution}\label{s_prp_slt}

\subsection{The Formulation of KANs}\label{ss_frm_kan}
The general formulation of a KAN can be expressed as:
\begin{align} \label{eq_f}
    f(\mathbb{X}) = \sum_{q=1}^{2\eta+1}{\Phi_{q} \Big( \sum_{\rho=1}^{\eta}{\phi_{q, \rho}(x_{\rho})} \Big)},
\end{align}
where $f$ is the continuous multivariate function being approximated, $\mathbb{X} = \{x_1, x_2, ..., x_{\eta}\}$ is the set of input variables, $\eta$ denotes the number of input variables, $\Phi_q$ ($q \in \{1, ..., 2\eta+1\}$) are continuous univariate functions forming the outer layer, and $\phi_{q,\rho}$ ($q \in \{1, ..., 2\eta+1\}$ and $\rho \in \{1, ..., \eta\}$) are continuous univariate functions forming the inner layer. This formulation establishes the fundamental structure of the KAN architecture. The inner functions $\phi_{q,\rho}$ constitute the first layer of the Kolmogorov-Arnold network, performing initial transformations on individual input variables. The outer functions $\Phi_q$ form the second layer, operating on the aggregated outputs of the corresponding inner functions.

As mentioned, the approximation process in a KAN is achieved through a systematic transformation and aggregation of input variables. The initial transformation is performed by the first layer, comprising inner functions $\phi_{q, \rho}$. This layer consists of $\eta \times (2\eta + 1)$ functions. Each function $\phi_{q, \rho}$ operates on a single input variable $x_{\rho}$, serving as a feature extractor to capture diverse aspects of the input. The outputs of these functions are subsequently aggregated for each $q$, yielding $2\eta + 1$ intermediate results. The original KAN paper proposes an implementation of the functions $\phi$ as described in \eqref{eq_phi}:
\begin{align} \label{eq_phi}
    {\phi}(x) &= \omega \times \big( b(x) + \text{spline}(x) \big) \notag\\
    &= \omega \times \Big( \frac{x}{1 + {e}^{-x}} + \sum_{i}^{}{c_{i} \times B_{i}(x)} \Big)
\end{align}
In this formulation, $\omega$ acts as a scaling factor, modulating the overall magnitude of the function's output. $b(x)$ represents a smooth approximation of the Rectified Linear Unit (ReLU) function, introducing non-linearity and facilitating the capture of complex data patterns. $\text{spline}(x)$ is a piecewise polynomial function, where $B_i(x)$ denote basis functions (typically B-spline basis functions), and $c_i$s are trainable coefficients or weights associated with each basis function.

The second layer comprises outer functions $\Phi_{q}$, which further process the intermediate results from the first layer. This layer consists of $2\eta + 1$ functions, where each function $\Phi_{q}$ takes as input the sum of the outputs from the corresponding $\phi_{q, \rho}$ functions of the first layer. These $\Phi_{q}$ functions perform additional transformations on the aggregated features, enabling more complex, non-linear mappings. The final output of the KAN is obtained by summing the outputs of all $\Phi_{q}$ functions. A generalized formulation of a KAN with $\mathcal{L} = 2\eta+1$ outer functions $\Phi_{q}$ can be expressed as follows:
\begin{align} \label{eq_kan}
    \text{KAN}(\mathbb{X}) = \big( {\Phi}_{\mathcal{L}-1} \circ {\Phi}_{\mathcal{L}-2} \circ \ldots \circ {\Phi}_{0} \big)({\mathbb{X}})
\end{align}
where $\circ$ denotes function composition. 

The universal approximation capability of KANs is derived from the flexibility in selecting appropriate $\phi_{q, \rho}$ and $\Phi_{q}$ functions. These functions can be optimized during the training process through the adjustment of trainable parameters (the $c_i$ coefficients in the spline functions). Once trained, these fully explainable functions can be used to directly produce outputs based on given inputs using basic mathematical operations, making KANs highly efficient in terms of inference cost and suitable particularly for real-time applications in resource-constrained environments.

\subsection{Dataset Generation}\label{ss_dt_gnr}
To address the $\alpha$-FPAP with KANs, the first step is to create a supervisory dataset, as represented in Algorithm \ref{alg_dataset_gen}. This algorithm systematically generates a diverse set of training examples, encompassing various network topologies and fairness parameters. The algorithm takes as input the system model parameters ($\Theta$) and the desired dataset size ($\mathcal{D}$). It iteratively generates random network topologies ($\mathcal{H}$) based on the system model parameters and solves the $\alpha$-FPAP for different values of $\alpha$ ranging from 0 to 0.9 in increments of 0.1. Each dataset entry corresponds to a specific $\alpha$ value and a network with randomly located UEs and BSs, featuring optimal transmission powers ($\mathbb{P}$) obtained from the Gurobi optimization solver \cite{gurobi}.

It's important to note that since the $\alpha$-FPAP is NP-hard, we employ practical approximations to make the method computationally feasible. Specifically, we utilize piece-wise linear approximations of the exponential function in the $\alpha$-fairness definition and the logarithmic function in Shannon rates. This approach enables us to extract near-optimal solutions to the problem for various inputs (i.e., UEs' path gains) and fairness-utilization trade-offs (via adjusting $\alpha$ values), thus creating a comprehensive training dataset for the framework. The algorithm continues to generate and solve instances until the desired dataset size $\mathcal{D}$ is reached. The resulting dataset $\mathcal{S}$ contains tuples of network topologies, $\alpha$ values, and their corresponding optimal transmission powers, providing a rich set of examples for training the KAN to approximate solutions to the $\alpha$-FPAP across different network configurations and fairness settings.

\subsection{Model Training}\label{ss_mdl_trn}
Following the dataset generation, the subsequent phase involves KAN training to address the $\alpha$-FPAP. We adopt a decentralized methodology wherein each BS is tasked with training its dedicated KAN to ascertain the transmit power of its associated UEs. As delineated in Algorithm \ref{alg_kan_training}, the training process commences with data preprocessing. For each BS $k$, we construct a new dataset $\mathcal{S}'_{k}$, derived from the original dataset $\mathcal{S}$. This dataset comprises input vectors $\mathbb{X}_t$, encompassing path gains $h_{i,k}$ for all UEs in the network topology $\mathcal{H}_t$ and the fairness parameter $\alpha_t$, along with target vectors $\mathbb{Y}_t$, consisting of optimal transmit powers $p_i$ for UEs associated with BS $k$. Next, the dataset is partitioned into training and test sets utilizing a ratio of $\beta$. We then initialize a KAN for each BS, configured with $\mathcal{B} \times \mathcal{N} + 1$ inputs (corresponding to the size of $\mathbb{X}_t$) and $\mathcal{N}/\mathcal{B}$ outputs (representing the size of $\mathbb{Y}_t$, assuming equal distribution of UEs among BSs). The KAN for each BS undergoes training using its respective training set and is subsequently evaluated using the test set.

\begin{algorithm}[b!]\label{alg_dataset_gen}
\caption{Dataset Generation}
\KwInput{System Model Parameters ($\Theta$), \\ \quad\qquad Dataset Size ($\mathcal{D}$)}
\KwOutput{Dataset of Optimal Transmission Powers ($\mathcal{S}$)}
Initialize $\mathcal{S} \gets \emptyset$ \\
$t \gets 0$ \\
\While{$|\mathcal{S}| < \mathcal{D}$}{
    Generate random network topology $\mathcal{H}_t$ based on $\Theta$ \\
    \For{$\alpha \in \{0, 0.1, 0.2, \ldots, 0.9\}$}{
        $\alpha_t \gets \alpha$ \\
        $\mathbb{P}_t \gets$ Solve the $\alpha$-FPAP for $(\mathcal{H}_t, \alpha_t)$ \\
        $\mathcal{S} \gets \mathcal{S} \cup \{t: (\mathcal{H}_t, \alpha_t, \mathbb{P}_t)\}$ \\
        $t \gets t + 1$ \\
        \If{$t = \mathcal{D}$}{
            \textbf{break}
        }
    }
}
\Return{$\mathcal{S}$}
\end{algorithm}

\begin{algorithm}[b!]\label{alg_kan_training}
\caption{KAN Training}
\KwInput{Dataset $\mathcal{S}$, Number of BSs $\mathcal{B}$, Training ratio $\beta$}
\KwOutput{Trained KAN for each BS}
\For{$k \in \mathbb{B}$}{
    Initialize $\mathcal{S}'_{k} \gets \emptyset$ \\
    \For{$t \in \{1, 2, \ldots, \mathcal{D}\}$}{
        extract $\{t: (\mathcal{H}_t, \alpha_t, \mathbb{P}_t)\}$ from $\mathcal{S}$ \\
        $\mathbb{X}_t \gets \{h_{i,k} | \forall i \in \mathbb{N} \text{ and } k \in \mathbb{B} \text{ of } \mathcal{H}_t\} \cup \{\alpha_t\}$ \\
        $\mathbb{Y}_t \gets \{p_i | \forall i \in \mathbb{N} \text{ and } b_i = k \text{ and } p_i \in \mathbb{P}_t\}$ \\
        $\mathcal{S}'_{k} \gets \mathcal{S}'_{k} \cup \{(\mathbb{X}_t, \mathbb{Y}_t)\}$
    }
    $\mathcal{S}^{train}_{k} \gets \beta \cdot |\mathcal{S}'_{k}|$ random samples from $\mathcal{S}'_{k}$ \\
    $\mathcal{S}^{test}_{k} \gets (1-\beta) \cdot |\mathcal{S}'_{k}|$ random samples from $\mathcal{S}'_{k}$ \\
    Initialize the KANs with shape $(\mathcal{B} \times \mathcal{N} + 1, \mathcal{N}/\mathcal{B})$\\
    Train the KAN of BS $k$ using $\mathcal{S}^{train}_{k}$ \\
    Evaluate the KAN of BS $k$ using $\mathcal{S}^{test}_{k}$ \\
}
\Return{Trained KAN for each BS}
\end{algorithm}

\section{Evaluation}\label{s_sim}

\subsection{Setup}
In this section, we conduct a numerical analysis of the proposed KAN-based solution using the system model parameters detailed in Table \ref{t_prm}. While other parameters may be selected arbitrarily, they must adhere to the logical framework established in Section \ref{s_prb_stt}. We explore two scenarios: first, evaluating the low inference cost promised by KANs, and second, assessing KAN's efficiency in solving the $\alpha$-FPAP.

\begin{table}[t!]
\caption{System Model Parameters.}
\begin{center}
\begin{tabular}{|c|c|}
\hline
\textbf{Parameter} & \textbf{Value} \\
\hline
Network area & 2D: $100 m \times 100 m$ \\
Noise power ($\sigma^2$) & $10^{-9}$ \\
Max transmit power ($\mathcal{P}_{\text{max}}$) & $1000 W$ \\
Min transmit power ($\mathcal{P}_{\text{min}}$) & $10 W$ \\
Path gain ($h_{i,j}$) & $\left( \| \textbf{loc}_i - \textbf{loc}_j \|_2 \right)^{-2}$ \\
$\alpha$ & $[0.1, 0.5, 0.9]$ \\
KAN implementation & Based on the design of Liu \textit{et al.} \cite{liu2024kan} \\
\hline
\end{tabular}
\label{t_prm}
\end{center}
\vspace{-15pt}
\end{table}

\subsection{Inference Cost}\label{ss_ntr}
\begin{figure}[!t]
    \centerline{\includegraphics[width=2in]{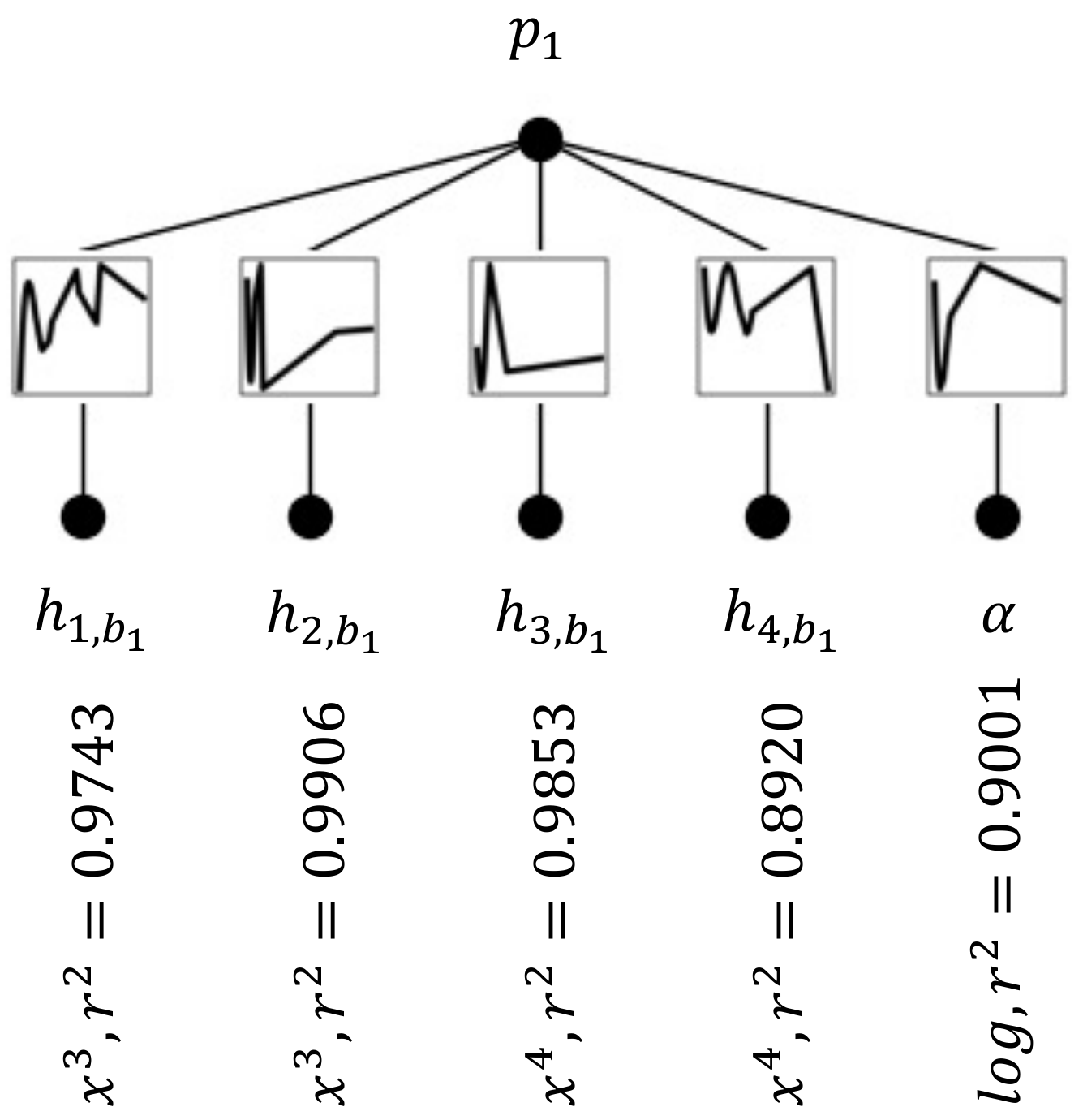}}
    \caption{The trained KAN for a network comprising four UEs and one BS and approximated functions for each element of the input vector $\mathbb{X} = \{ h_{1, b_1}, h_{2, b_1}, h_{3, b_1}, h_{4, b_1}, \alpha \}$ to compute $p_1$. It is important to note that $r^2$ is the coefficient of determination, indicating the accuracy with which the symbolic function approximates the underlying data for each element.}
    \label{fig3}
\end{figure}

To evaluate the inference cost of KANs, we consider a scenario involving four UEs and one BS. After training the KAN for 100 rounds using Algorithm \ref{alg_kan_training}, we approximate the function applied to each element of the input vector $\mathbb{X}$ to generate the corresponding element of the target vector $\mathbb{Y}$. Figure \ref{fig3} depicts the functions relating each element of $\mathbb{X}$ to the calculation of the transmit power of UE 1 ($p_1$) after pruning. The relaxed general function for computing $p_1$ based on $\mathbb{X}$ is given by:
\begin{equation}
\begin{aligned}
    \text{KAN}(\mathbb{X})[p_1] \simeq & -8031 \times (0.06 - x_1)^3 + 561 \times (0.09 - x_2)^3 \\
    & + 6440 \times (0.1 - x_3)^4 + 9121 \times (0.09 - x_4)^4 \\
    & - 0.06 \times \log(0.4 \times x_5) - 0.19 \notag
\end{aligned}
\end{equation}
This explainable structure enables us to understand the contribution of each input feature to the output, allowing decision-making through straightforward arithmetic operations. Consequently, when the number of UEs or parameter $\alpha$ changes in our scenario, the transmit powers can be determined with significantly reduced inference costs, in contrast to many other deep machine learning techniques that necessitate extensive computational resources due to their complexity.

\subsection{fairness}\label{ss_ffc}
\begin{figure}[!t]
    \centerline{\includegraphics[width=3in]{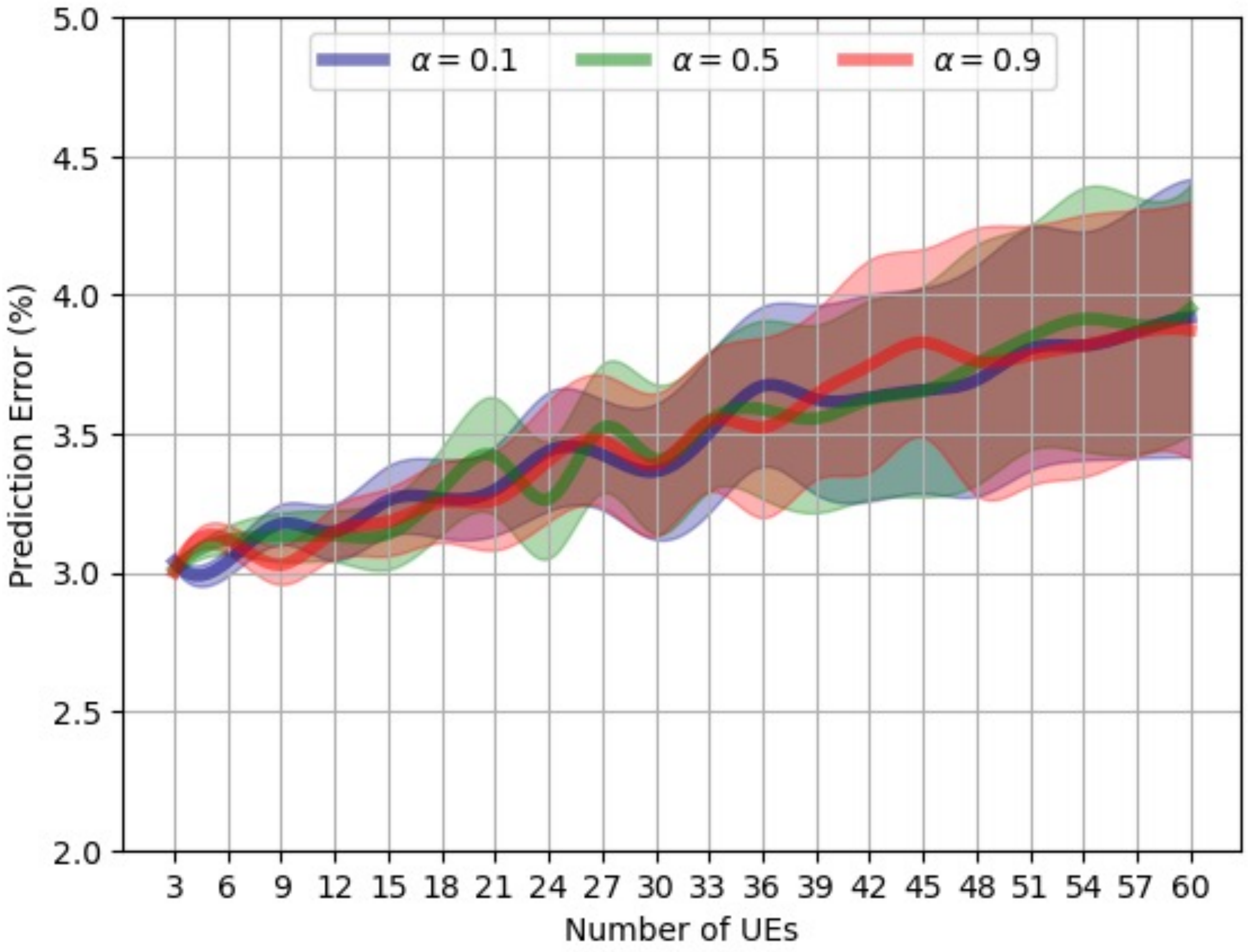}}
    \caption{The prediction error (\%) for the proposed KAN solution in a network with three BSs and varying numbers of UEs (3-60), shown for different $\alpha$ values (0.1, 0.5, 0.9). The solid lines represent the average of multiple simulation results, while the shaded areas encompass the individual data points from each simulation run, illustrating the range and distribution of outcomes.}
    \label{fig4}
\end{figure}

To investigate the efficiency of the proposed solution in terms of fairness, we consider a scenario involving three BSs and the number of UEs varying from 3 to 60. After training the KAN for 10000 rounds, the results for different $\alpha$ values are illustrated in Fig. \ref{fig4}. The figure demonstrates that the prediction error is consistently low across different values of $\alpha$, indicating the robustness and effectiveness of the KAN in allocating transmit powers. When the number of UEs is small, the error is around 3\%, and as the number of UEs increases to 60, the error only grows to about 4\%. Considering the NP-hard nature of the problem, which typically results in exponential growth in complexity as the problem size increases, this modest increase in error is remarkably small. This demonstrates that the KAN remains highly efficient even as the network scales up significantly.

\section{Conclusion}\label{s_con}
This paper addresses the fairness-utilization trade-off in wireless networks by maximizing $\alpha$-fairness of UE data rates through transmit power allocation. First, we formulated the problem as a non-linear program and proved its NP-hardness by reducing the MISP to it. Then, to achieve low inference costs, we proposed a method based on the Kolmogorov-Arnold representation theorem, detailing data generation and decentralized KAN training algorithms. Extensive simulations demonstrated the KAN's high efficiency across significant network size increases and various $\alpha$ values, allocating transmit powers with negligible computational cost due to its explainability.

For future research directions, we propose to extend the application of KAN-based decision-making to multiple-access control problems. This extension would encompass not only the management of UE transmit power but also the regulation of their admission to shared spectrum resources. Such an approach would build upon our previous work on multiple access in semantic-aware \cite{mazandarani2024semantic} and continual reinforcement learning \cite{10271915} scenarios within the Metaverse context. Furthermore, we intend to explore the integration of KAN-based power allocation techniques with resource allocation strategies in other domains. This includes investigating applications in wired network infrastructures and distributed computing environments \cite{shokrnezhad2024orient, 10436905, 10207694}.

\bibliographystyle{IEEEtran}
\bibliography{IEEEabrv,Bibliography}

\end{document}